\documentclass[12pt]{article}
\usepackage{amssymb}

\usepackage{pifont}
\usepackage{mathrsfs}
\usepackage{amsfonts}
\usepackage{color}
\usepackage{footmisc}
\usepackage[numbers,sort&compress]{natbib}
\usepackage[bookmarksopen=true, bookmarksopenlevel=\maxdimen,
bookmarksnumbered=true,colorlinks,linkcolor=blue,
citecolor=blue,urlcolor=blue]{hyperref}

\newcommand{\omits}[1]{}

\definecolor{darkgreen}{rgb}{0.0,0.6,0.0}
\definecolor{lightgrey}{rgb}{0.7,0.7,0.7}

\begin{document}
\begin{center}
{\bf \LARGE Cosmological meaning of the\\ \bigskip
gravitational gauge group}

\bigskip

\bigskip

{\Large Jia-An Lu$^{a}$\footnote{Email: ljagdgz@163.com}}

\bigskip

$^a$ School of Physics and Engineering, Sun Yat-sen
University,\\ Guangzhou 510275, China

\setcounter{footnote}{-1}
\footnote{The final publication is available at Springer via
\href{http://dx.doi.org/10.1007/s10714-014-1780-5}
{http://dx.doi.org/10.1007/s10714-014-1780-5}.}

\begin{abstract}
It is shown that among the $R+\beta S^{abc}S_{abc}$ models,
only the one with $\beta=1/2$ has nonvanishing torsion effect
in the Robertson--Walker universe filled with a spin fluid,
where $S_{abc}$ denotes torsion. Moreover, the torsion effect
in that model is found to be able to replace the big-bang singularity
by a big bounce. Furthermore, we find that the model can be obtained
under a Kaluza--Klein-like ansatz, by assuming that the gravitational
gauge group is the de Sitter group.
\end{abstract}
\end{center}



\section{Introduction}

It has been pointed out that there are three kinds of special relativity
(SR) \cite{BdS,dSSR,SR-wick,PoR-SR}
and gravity should be based on the localization of a SR with full symmetry
\cite{PoI,Guo07,SR-Gravity}. It is a motivation for the study of the
Poincar\'e, de Sitter (dS) or Anti-de Sitter (AdS) gauge theory of gravity
\cite{Hehl76,Guo76,Stelle,Grignani,Lu13}, where the Riemann--Cartan (RC) geometry
with nontrivial metric and torsion is introduced to realize the
corresponding gauge symmetry.

The Kaluza--Klein-like (KK-like) models \cite{Mansouri,Guo79,Guo79-dS,Lu13}
of the gauge theory of gravity are some models in which the Lagrangians
are constructed from the scalar curvatures of
some fiber bundles, just like the KK theory (for example, see Refs. \cite{Cho,Percacci}).
The scalar curvature of the bundle is given by the RC geometries of
both the spacetime and the typical fiber.
In such models, one can see what effects the geometry of the typical fiber,
which is determined by the gravitational gauge group, would have on the gravitational dynamics.
In other words, one can see what effects the gravitational
gauge group would have on the gravitational dynamics.

In the previous KK-like models, it is shown that the fiber geometries
produce some quadratic torsion terms \cite{Lu13}, or together with some quadratic
curvature terms \cite{Mansouri,Guo79}, in the gravitational Lagrangian.
In these models, however, almost the same dynamics are obtained, when
different gauge groups are used. In order to find the effects caused by the
difference of the gauge groups, we introduce a coupling constant $\alpha$
between the spacetime and the typical fiber, and construct the corresponding KK-like
models in this paper. It is found that the dS group can
decide the value of $\alpha$ while the Poincar\'e group cannot, by the requirement
that the no-gravity spacetime is a vacuum solution.

Moreover, in this paper, it is shown that among the $R+\beta S^{abc}S_{abc}$ models,
only the one with $\beta=1/2$ has nonzero torsion effect in the homogeneous
and isotropic universe filled with a spin fluid, where $S_{abc}$ denotes
torsion. Also, the torsion effect of the model is found to be able to avert
the big-bang singularity of the homogeneous and isotropic universe.
On the other hand, somewhat inconceivably, we find that the model is just
the aforementioned KK-like model determined by the dS group.

The paper is organized as follows. In section 2, some new KK-like models
are constructed by introducing the coupling constant $\alpha$.
In section 3, it is shown that among a class of models, the KK-like model
determined by the dS group is the only one that may avert the big-bang singularity
under certain assumptions. Finally, we give some remarks in the last section.

\section{Kaluza\texorpdfstring{--}{-}Klein-like models}

Let ${\cal M}$ be the spacetime manifold, with the metric $g_{ab}$,
and the metric-compatible derivative operator $\nabla_a$,
where $a,b$ are the abstract indices \cite{Wald,Liang}.
There exist tensor fields $S^{c}{}_{ab}$ and $R^c{}_{dab}$, such that
for any function $f$ and 1-form field $\omega_{a}$ on ${\cal M}$,
\begin{equation}\label{torsion}
(\nabla_{a}\nabla_{b}-\nabla_{b}\nabla_{a})f=
-S^{c}{}_{ab}\nabla_{c}f,
\end{equation}
\begin{equation}\label{curvature}
(\nabla_{a}\nabla_{b}-\nabla_{b}\nabla_{a})\omega_{d}
=-R^c{}_{dab}\,\omega_{c}-S^{c}{}_{ab}\nabla_{c}\,\omega_{d}.
\end{equation}
The tensor fields $S^{c}{}_{ab}$ and $R^c{}_{dab}$ are called the
torsion and curvature tensor fields of $\nabla_a$, respectively.
Moreover, the Ricci tensor field and the scalar curvature are
defined as $R_{ab}=R^c{}_{acb}$ and $R=g^{ab}R_{ab}$, respectively.
Suppose that $\{e_{\alpha}{}^a\}$ is an orthonormal frame field,
where $\alpha=0,1,2,3$. The signature is chosen such
that the metric components in $\{e_{\alpha}{}^a\}$ are equal to
$\eta_{\alpha\beta}=$ diag$(-1,1,1,1)$. The connection 1-form fields of
$\nabla_a$ in $\{e_{\alpha}{}^a\}$ are defined as
\begin{equation}
\Gamma^{\alpha}{}_{\beta
a}=e^{\alpha}{}_{b}\nabla_{a}e_{\beta}{}^{b},
\end{equation}
where $\{e^{\alpha}{}_{b}\}$ is the dual frame field
of $\{e_{\alpha}{}^a\}$. The torsion and curvature 2-form fields of $\nabla_a$
in $\{e_{\alpha}{}^a\}$ are defined as
$S^{\alpha}{}_{ab}=S^{c}{}_{ab}\,e^{\alpha}{}_{c}$ and
$R^{\alpha}{}_{\beta ab}=R^c{}_{dab}\,e^{\alpha}{}_{c}\,e_{\beta}{}^{d}$, respectively.
Then Eqs. (\ref{torsion}) and (\ref{curvature}) imply
\begin{equation}
S^{\alpha}{}_{ab}=d_a e^{\alpha}{}_b+\Gamma^{\alpha}{}_{\beta
a}\wedge e^{\beta}{}_{b},
\end{equation}
\begin{equation}
R^{\alpha}{}_{\beta ab}=d_a\Gamma^{\alpha}{}_{\beta b}+\Gamma^{\alpha}{}_{\gamma
a}\wedge \Gamma^{\gamma}{}_{\beta b},
\end{equation}
where $d_a$ is the exterior differentiation operator of ${\cal M}$.

The gauge-invariant expressions for the metric, torsion and curvature
tensor fields of the spacetime are as follows \cite{Guo76,Stelle,Grignani,Lu13}:
\begin{equation}\label{metricxi}
g_{ab}=\eta_{AB}(D_{a}\xi^{A})(D_{b}\xi^{B}),
\end{equation}
\begin{equation}\label{torsionxi}
S_{cab}=\mathcal{F}_{ABab}(D_{c}\xi^{A})\xi^{B},
\end{equation}
\begin{equation}\label{R-F}
R_{cdab}-(2\epsilon/l^{2})g_{a[c}g_{d]b}=\mathcal
{F}_{ABab}(D_{c}\xi^{A})(D_{d}\xi^{B}),
\end{equation}
where $S_{cab}=g_{cd}S^d{}_{ab}$, $R_{cdab}=g_{ce}R^e{}_{dab}$,
$\epsilon=0$ or 1, $l$ is a constant with the dimension of length,
$A,B=0,1,2,3,4$, $\eta_{AB}={\rm diag}(-1,1,1,1,1)$,
$D_a\xi^A=d_a\xi^A+\Omega^A{}_{Ba}\xi^B$, $\Omega^A{}_{Ba}$
is the Ehresmann connection of a principal bundle ${\cal P}_G$,
${\cal F}_{ABab}=\eta_{AC}{\cal F}^C{}_{Bab}$, ${\cal F}^C{}_{Bab}=
d_a\Omega^C{}_{Bb}+\Omega^C{}_{Da}\wedge\Omega^D{}_{Bb}$,
$\xi^A=\xi^A(x)$ are coordinates of a global section $\phi$ in a fiber
bundle ${\cal Q}_F$ as a subbundle of ${\cal Q}_{M_5}$, both of ${\cal Q}_F$
and ${\cal Q}_{M_5}$ are associated to ${\cal P}_G$, and the typical
fiber $M_5$ of ${\cal Q}_{M_5}$ is the 5-dimensional (5d) Minkowski space.
When the structure
group $G$ of ${\cal P}_G$ is the dS group $S_{10}$, the typical fiber
$F$ of ${\cal Q}_F$ is a 4d dS space $D_4$ with radius $l$,
and $\epsilon=1$. When the structure group $G$ is the Poincar\'e group $P_{10}$,
the typical fiber $F$ is the 4d Minkowski space $M_4$, and $\epsilon=0$.

Note that $\xi^A$ can be interpreted as the localized 5d Minkowski coordinates,
for the reason that the metric field in $F$ can be expressed as
$g_{ab}=\eta_{AB}(d_a\xi^A)(d_b\xi^B)$, where $\xi^A$ are the 5d Minkowski coordinates
on $F$ as a hypersurface of $M_5$. The localized 5d Minkowski coordinates $\xi^A$
are related to the local inertial coordinates $y^\mu$ on the spacetime ${\cal M}$,
by $y^\mu=\xi^\mu$ when $G=P_{10}$, or
$y^\mu=\sqrt{\sigma}\xi^\mu$ when $G=S_{10}$ and $\xi^4\neq0$,
where $\sigma=(l/\xi^4)^2$, and $\mu=0,1,2,3$ \cite{Grignani,dSSR,Guo76}.

The no-gravity spacetime can be defined as a spacetime where ${\cal F}^A{}_{Bab}=0$.
The gauge-invariant expressions (\ref{torsionxi}) and (\ref{R-F}) imply that
the no-gravity spacetime is locally identical with the typical fiber $F$.
It is natural to require that the no-gravity spacetime is a vacuum solution of the
gravitational field equations.

The KK-like ansatz we are going to propose is as follows: the gravitational
Lagrangian is given by \cite{Lu13}
\begin{equation}\label{Lg}
\mathscr{L}_g=\phi^*\overline{R},
\end{equation}
where $*$ denotes a pullback, and $\overline{R}$ is the scalar curvature
of ${\cal Q}_F$, which is determined by the bundle metric
\begin{equation}\label{gbar}
\overline{g}=g_{\mu\nu}dx^{\mu}\otimes
dx^{\nu}+\alpha\eta_{AB}\theta^{A}\otimes\theta^{B}
\end{equation}
and the bundle torsion
\begin{equation}
\overline{S}=S^\sigma{}_{\mu\nu}X_\sigma\otimes dx^\mu\otimes dx^\nu,
\end{equation}
where $d$ is the exterior differentiation operator of ${\cal Q}_F$,
$\otimes$ stands for a tensor product, $\alpha\neq0$ is a dimensionless
coupling constant between ${\cal M}$ and $F$,
$\theta^A=d\xi^A+\Omega^A{}_{B\mu}\xi^Bdx^\mu$,
$X_\sigma=\partial_\sigma-\Omega^A{}_{B\sigma}\xi^B\partial_A$,
$x^\mu$ are coordinates on ${\cal Q}_F$ and are induced from a coordinate
system $\{x^\mu\}$ on ${\cal M}$, $\partial_\sigma=\partial/\partial x^\sigma$,
$\xi^A$ (here differ from $\xi^A(x)$) are coordinates on ${\cal Q}_F$ as
a subbundle of ${\cal Q}_{M_5}$, and are induced from a 5d Minkowski coordinate
system of $M_5$, $\partial_A=\partial/\partial\xi^A$,
$g_{\mu\nu}$, $S^\sigma{}_{\mu\nu}$ and $\Omega^A{}_{B\mu}$ are components of
$g_{ab}$, $S^c{}_{ab}$ and $\Omega^A{}_{Ba}$ in $\{x^\mu\}$.
Notice that $\alpha=1$ in Ref. \cite{Lu13}, while it is to be determined here.
After some calculations, it can be checked that
\begin{equation}
\phi^*\overline{R}=R+4\epsilon\Lambda/\alpha-(1/4)\alpha|S|^2,
\end{equation}
where $\Lambda=3/l^2$ is the cosmological constant of $D_4$, and
$|S|^2=S^{abc}S_{abc}$. When $G=S_{10}$, $\epsilon=1$, and, to ensure that the
no-gravity spacetime is a vacuum solution, $4\Lambda/\alpha$ should be
equal to $-2\Lambda$. As a result, $\alpha=-2$, and the gravitational Lagrangian is
\begin{equation}\label{Lagrangian}
\mathscr{L}_g=\phi^{*}\overline{R}=R-2\Lambda+(1/2)|S|^2.
\end{equation}
When $G=P_{10}$, $\epsilon=0$, and so the requirement that the no-gravity
spacetime is a vacuum solution gives no constraint on $\alpha$.
Consequently, the gravitational Lagrangian is
\begin{equation}\label{Lagrangianp}
\mathscr{L}_g=\phi^{*}\overline{R}=R-(1/4)\alpha|S|^2.
\end{equation}
It can be seen from Eqs. (\ref{Lagrangian}) and (\ref{Lagrangianp})
that different gravitational gauge groups result in different gravitational dynamics.

The following class of Lagrangians contains both of Eqs. (\ref{Lagrangian}) and
(\ref{Lagrangianp}) as special cases:
\begin{equation}\label{Lagrangian2}
\mathscr{L}_g=R-2\epsilon\Lambda+\beta|S|^2,
\end{equation}
where $\beta$ is a dimensionless parameter. The gravitational field equations
of the above models are as follows:
\begin{eqnarray}\label{1stEq}
R_{ba}-\frac1 2Rg_{ab}+\epsilon\Lambda g_{ab}-2\beta\nabla_{c}S_{ab}{}^{c}
-\beta S_{acd}T_{b}{}^{cd}-\frac1 2\beta|S|^2g_{ab}
+2\beta S^{cd}{}_aS_{cdb}=\frac{1}{2\kappa}\Sigma_{ab},
\end{eqnarray}
\begin{equation}\label{2ndEq}
 T^{a}{}_{bc}-4\beta S_{[bc]}{}^{a}=-\frac{1}{\kappa}\tau_{bc}{}^{a},
\end{equation}
where $T^a{}_{bc}=S^a{}_{bc}+2\delta^a{}_{[b}S_{c]}$,
$\delta^a{}_b$ is the Kronecker delta, $S_c=S^b{}_{cb}$,
$\kappa$ is the gravitational coupling constant,
$\Sigma_{ab}=g_{bc}(\delta S_m/\delta e^\alpha{}_c)e^\alpha{}_a$
is the canonical energy-momentum tensor,
$\tau_{bc}{}^a=(\delta S_m/\delta\Gamma^\alpha{}_{\beta a})e^\alpha{}_be_{\beta c}$
is the spin tensor, where $e_{\beta c}=\eta_{\alpha\beta}e^\alpha{}_c$,
and $S_m$ is the action of the matter fields.
Note that the cases with $\beta=1,1/4$ or $-1/2$ are problematic
because they can only describe the matter fields with $\tau_b\equiv\tau_{bc}{}^c=0$,
$\tau_{[abc]}=0$, or $g_{a[b}\tau_{c]}-\tau_{a[bc]}+\tau_{bca}=0$, respectively.
For the cases with $\beta\neq1,1/4,-1/2$, the solution
of Eq. (\ref{2ndEq}) is
\begin{equation}\label{tor-spin}
S_{abc}=(8\beta^2+2\beta-1)^{-1}[-2(4\beta-1)g_{a[b}S_{c]}
-(2\beta-1)(1/\kappa)\tau_{bca}+(4\beta/\kappa)\tau_{a[bc]}],
\end{equation}
where $S_c=[1/2\kappa(1-\beta)]\tau_c$.
The energy-momentum tensor $T_{ab}=-2\delta S_M/\delta g^{ab}$ is related to
$\Sigma_{ab}$ and $\tau_{bc}{}^a$ by
\begin{eqnarray}\label{T-Sigma}
T_{ab}&=&\Sigma_{ab}+(\nabla_c+S_c)(\tau_{ab}{}^c-\tau_a{}^c{}_b+\tau^c{}_{ba})\nonumber\\
&=&\Sigma_{(ab)}+2(\nabla_c+S_c)\tau^c{}_{(ab)}.
\end{eqnarray}
Hence, if the spin tensor is equal to zero, then $S^c{}_{ab}=0$,
$\Sigma_{ab}=T_{ab}$, and Eq. (\ref{1stEq}) reduces to the Einstein field
equation when $1/2\kappa=8\pi$, i.e., $\kappa=1/16\pi$. Generally, Eq. (\ref{1stEq})
can be expressed as an Einstein-like equation:
\begin{equation}\label{Elike}
\mathring R_{ab}-\frac1 2\mathring Rg_{ab}+\epsilon\Lambda g_{ab}
=\frac1{2\kappa}(T_{\rm eff})_{ab},
\end{equation}
where $\mathring R=g^{ab}\mathring R_{ab}$, $\mathring R_{ab}=\mathring R^c{}_{acb}$,
$\mathring R^c{}_{dab}$ is the torsion-free curvature tensor,
and $(T_{\rm eff})_{ab}$ is the effective energy-momentum tensor which satisfies
\begin{eqnarray}\label{Teff}
\frac1{2\kappa}(T_{\rm eff})_{ab}=\frac1{2\kappa}T_{ab}
-\frac1\kappa(\frac12S_{(a}{}^{cd}\tau_{|cd|b)}+K^d{}_{(b|c|}\tau^c{}_{a)d})
+\frac12(1+2\beta)S^{cd}{}_{(a}S_{b)cd}\nonumber\\
-S_{(ab)}{}^cS_c-\frac14(T_{cde}K^{dec}-2\beta|S|^2)g_{ab}
-\frac14S_{acd}S_b{}^{cd}-2\beta S_{cda}S^{[cd]}{}_b,
\end{eqnarray}
where $K^d{}_{bc}=(1/2)(S^d{}_{bc}+S_{bc}{}^d+S_{cb}{}^d)$
is the contorsion tensor.

\section{Cosmological meaning}

Let us assume that the matter fields in the universe
can be described by a spin fluid \cite{Weyssenhoff,Kuchowicz1}
with the energy-momentum tensor and spin tensor being
\begin{equation}\label{T}
T_{ab}=\rho U_aU_b+p(g_{ab}+U_aU_b),
\end{equation}
\begin{equation}\label{tau}
\tau_{bc}{}^a=\tau_{bc}U^a,
\end{equation}
where $\rho$ is the rest energy density, $p$ is the hydrostatic pressure,
$U^a$ is the four-velocity
of the fluid particles, and $\tau_{bc}$ is the spin density 2-form which
satisfies $\tau_{bc}U^c=0$. Substitution of
Eq. (\ref{tau}) into Eq. (\ref{Teff}) yields
\begin{eqnarray}\label{Teff2}
\frac1{2\kappa}(T_{\rm eff})_{ab}=\frac1{2\kappa}T_{ab}
+\frac1{\kappa^2}(8\beta^2+2\beta-1)^{-2}
[(\beta-\frac12)(8\beta^2+2\beta-1)\tau_{ac}\tau_b{}^c\nonumber\\
+(-4\beta^3-3\beta^2+\frac14)2s^2U_aU_b
+(-6\beta^3-\frac12\beta^2+\beta-\frac18)2s^2g_{ab}],
\end{eqnarray}
where $s^2=\tau_{bc}\tau^{bc}/2$ is the spin density squire.
Furthermore, the metric field of the universe
is supposed to be a Robertson--Walker (RW) metric with the line element
\begin{equation}\label{RW}
ds^2=-dt^2+a^2(t)[dr^2+r^2(d\theta^2+\sin^2\theta d\varphi^2)],
\end{equation}
where $a(t)>0$ is called the scale factor, and $t$ is the cosmic time with
$(\partial/\partial t)^a=U^a$. According to Eq. (\ref{RW}), the left-hand side of
the Einstein-like equation (\ref{Elike}) is diagonal in the coordinate system
$\{t,r,\theta,\varphi\}$, then the effective energy-momentum tensor $(T_{\rm eff})_{ab}$
should be diagonal, and so the term containing $\tau_{ac}\tau_b{}^c$ in Eq.
(\ref{Teff2}) should be diagonal, which implies that $\beta=1/2$ or $\tau_{bc}=0$.
Note that substitution of $\beta=1/2$ into Eq. (\ref{Lagrangian2})
leads to Eq. (\ref{Lagrangian}). In other words, under the KK-like ansatz (\ref{Lg}),
the dS group can choose the only one model that is nontrivial in the RW universe
filled with a spin fluid (\ref{T})(\ref{tau}), among the $R-2\epsilon\Lambda+\beta|S|^2$
models, while the Poincar\'e group cannot.

It should be remarked that the torsion field given by Eqs. (\ref{tor-spin})
and (\ref{tau}) is not homogenous and isotropic in the usual sense.
In other words, the Lie derivatives of the torsion field along the Killing
vector fields which represent the homogeneous and isotropic properties
are not equal to zero. However, we may still think that
the torsion field is homogeneous and isotropic for the reason that
it is determined by the spin tensor by Eq. (\ref{tor-spin}), and the
spin tensor is homogeneous in the sense that the spin density $s$ is
a function of the cosmic time, and isotropic in the sense that the spin orientations
given by the spin density 2-form $\tau_{bc}$ are random at the cosmic scale.

As a matter of fact, each point of the cosmic space corresponds to a region
which is small relative to the cosmic space. If the spin orientations
are random in such a region, the spin density 2-form should be equal to zero
at that point corresponding to this region. In order to obtain a nonzero spin
density 2-form, we assume that the spin orientations are aligned in each of such
regions.

For the KK-like model (\ref{Lagrangian}), Eq. (\ref{Teff2}) becomes
\begin{equation}
(T_{\rm eff})_{ab}=T_{ab}-\frac1{2\kappa}s^2U_aU_b
-\frac1{2\kappa}s^2(g_{ab}+U_aU_b).
\end{equation}
Substitution of $\kappa=1/16\pi$ and Eq. (\ref{T}) into the above equation
leads to
\begin{equation}\label{Teff3}
(T_{\rm eff})_{ab}=(\rho+\rho_S)U_aU_b+(p+p_S)(g_{ab}+U_aU_b),
\end{equation}
where $\rho_S=p_S=-8\pi s^2$. The early universe is filled with
elementary bosons and fermions. In order to preserve the gauge symmetries,
we assume that the spin of any massless boson is not coupled to torsion \cite{Hehl76}.
Furthermore, the spin orientations of the massive bosons and fermions
are assumed to be locally aligned, and so $s=\hbar(n_{mb}+n_f/2)$, where $n_{mb}$
and $n_f$ are the number densities of the spin-1 massive bosons and fermions,
respectively. Moreover, it holds \cite{Kolb} that $\rho=3p=(\pi^2/30)g_*T^4$,
$n_{mb}=[\zeta(3)/\pi^2]g_{mb}T^3$ and $n_f=[3\zeta(3)/4\pi^2]g_fT^3$ when the
temperature $T\gtrsim300$ GeV$=3.480\times10^{15}$ K, where $\zeta(3)\approx1.202$ is the
Riemann zeta function of 3, $g_*=g_b+(7/8)g_f$, and, $g_b,\ g_{mb}$, and $g_f$ are
the sums of the numbers of interacting spin states for each type of bosons, spin-1
massive bosons and fermions, respectively. For all the known elementary particles, including
the Higgs bosons, $g_b=28$, $g_{mb}=9$, and $g_f=90$ \cite{Rich}.
Substituting the above relations into
\begin{equation}
\dot\rho+\dot\rho_S+(3\dot a/a)(\rho+p+2\rho_S)=0,
\end{equation}
which is the conservation law given by the Einstein-like equation (\ref{Elike})
and the RW metric, where $\cdot$ denotes the derivative with respect to $t$,
leads to $\dot T/T=-\dot a/a$ when $2\rho+3\rho_S\neq0$, with the solution
\begin{equation}\label{a-T}
a=CT^{-1},
\end{equation}
which is the same as the standard cosmology, where $C$ is a positive constant.
Furthermore, the Friedmann-like equation obtained from the Einstein-like equation
(\ref{Elike}) and the RW metric is as follows:
\begin{equation}\label{Flike}
(\dot{a}/a)^2=(8\pi/3)(\rho+\rho_S+\rho_\Lambda),
\end{equation}
where $\rho_\Lambda=\Lambda/8\pi$ can be neglected in the early universe.
Consequently, the minimum of the scale factor is approximately given by
$\rho+\rho_S=0$, with the solution $T=T_b\equiv\sqrt{A/B}$ (in the units with $\hbar=c=
G=k=1$), where $A=(\pi^2/30)g_*$, $B=2[\zeta(3)g_n]^2/\pi^3$, where
$g_n=2g_{mb}+3g_f/4$. In the international units,
\begin{equation}\label{Tb}
T_b=0.2270\times T_p=3.217\times10^{31}{\rm\ K},
\end{equation}
where $T_p=\sqrt{\hbar c^5/Gk^2}$ is the Planck temperature. In conclusion,
the torsion effect may replace the big-bang singularity by a big bounce
with the temperature (\ref{Tb}).

The local alignments of the spin orientations may be caused by the cosmic magnetic
fields \cite{Trautman}, which become weaker when the universe expands, and so
the spin orientations would eventually become random, and the torsion effect would
vanish.

Note that there is a similar discussion on the early universe \cite{Poplawski},
where a different spin fluid \cite{Hehl74} is used,
which satisfies Eq. (\ref{tau}) and
\begin{equation}\label{Sigma}
\Sigma_{ab}=W_aU_b+p(g_{ab}+U_aU_b),
\end{equation}
where $W_a$ is the momentum density 1-form. Utilizing the spin
conservation law $(\nabla_c+S_c)\tau_{ab}{}^c=-\Sigma_{[ab]}$ and Eq. (\ref{tau}),
one can find that
\begin{equation}\label{Sigma2}
\Sigma_{ab}=\rho U_aU_b+p(g_{ab}+U_aU_b)+2U_bU^cU^d\mathring\nabla_c\tau_{ad},
\end{equation}
where $\rho=-W_aU^a$ is the rest energy density. Substitution of Eqs. (\ref{Sigma2})
and (\ref{tor-spin}) into Eq. (\ref{T-Sigma}) leads to
\begin{eqnarray}\label{T2}
T_{ab}=\rho U_aU_b+p(g_{ab}+U_aU_b)+2\mathring\nabla_c\tau^c{}_{(ab)}
+2U^cU^dU_{(b}\mathring\nabla_{|c|}\tau_{a)d}\nonumber\\
+(2\beta+1)^{-1}(4\beta-1)^{-1}(1-2\beta)\frac1\kappa(2s^2U_aU_b+\tau_a{}^c\tau_{bc}).
\end{eqnarray}
Substituting the above equation into Eq. (\ref{Teff2}) yields
\begin{eqnarray}\label{Teff4}
(T_{\rm eff})_{ab}=(\rho+\rho_S)U_aU_b+(p+p_S)(g_{ab}+U_aU_b)\nonumber\\
+2\mathring\nabla_c\tau^c{}_{(ab)}+2U^cU^dU_{(b}\mathring\nabla_{|c|}\tau_{a)d},
\end{eqnarray}
where
\begin{eqnarray}
\rho_S=p_S=\frac4\kappa(2\beta+1)^{-2}(4\beta-1)^{-2}
(-6\beta^3-\frac12\beta^2+\beta-\frac18)s^2.
\end{eqnarray}
For the RW universe, the last term of Eq. (\ref{Teff4}) is equal to zero,
and the nonzero components of the term $2\mathring\nabla_c\tau^c{}_{(ab)}$
in Eq. (\ref{Teff4}) are:
\begin{equation}\label{curl}
2\mathring\nabla_c\tau^c{}_{(0i)}=\hat\nabla^j\hat\tau_{ij}
=\hat\varepsilon_{ijk}\hat\nabla^j(^*\hat\tau)^k\equiv[{\rm curl}(^*\hat\tau)]_i,
\end{equation}
where $\hat\nabla_i$ is the $\hat g_{ij}$-compatible
torsion-free derivative operator, $\hat g_{ij}$ is the confinement of $g_{ab}$
on the cosmic space $\Sigma_t$, $\hat\tau_{ij}$ is the confinement of $\tau_{bc}$
on $\Sigma_t$, $\hat\varepsilon_{ijk}$ is the Levi-Civita symbol,
$(^*\hat\tau)_k=\hat\tau^{ij}\hat\varepsilon_{ijk}/2$ is the spin density 1-form, and,
$i,j,k=1,2,3$ and are raised by the inverse of $\hat g_{ij}$. Since the off-diagonal
term of $(T_{\rm eff})_{ab}$ should be equal to zero, Eq. (\ref{curl}) should be
identical to zero, which may be inconsistent with the cosmic-scale randomness assumption
on the spin orientations. In Refs. \cite{Hehl74,Poplawski},
the term $2\mathring\nabla_c\tau^c{}_{(ab)}$ in Eq. (\ref{Teff4}) with $\beta=0$
is discarded by an averaging argument: assume that the microscopic spin orientations
are random, then the spin tensor and its derivative vanish after macroscopic averaging.
However, to my opinion, the averaging procedure is ambiguous.
If we average the spin tensor before substituting it into the gravitational field
equations, no torsion effect can survive. If the spin tensor is averaged after
being substituted into the gravitational field equations, the gravitational field
equations should be averaged at the same time, which is an ill-defined concept in
a classical theory of gravity.

\section{Remarks}

We find that under the KK-like ansatz (\ref{Lg}), the dS group can choose
the only one model that may avert the big-bang singularity of the RW universe
filled with a spin fluid (\ref{T})(\ref{tau}) by torsion, among the
class of models (\ref{Lagrangian2}), while the Poincar\'e group cannot.
Moreover, the dS group also presents an explanation for the cosmological
constant: it is related to the radius $l$ of the no-gravity spacetime by
$\Lambda=3/l^2$. It seems that a regular, accelerating universe favours the
dS group as the gravitational gauge group.

\section*{Acknowledgments}
I would like to thank the late Prof. H.-Y. Guo, and Profs. C.-G.
Huang, Z.-B. Li, X.-P. Zhu, X.-W. Liu, S.-D. Liang and T. Harko for their help
and some useful discussions. I would also like to thank Prof. Y.
Obukhov for the discussion on the spin fluid models.

\end{document}